\documentclass[namedreferences]{solarphysics}
\usepackage[hyperref,optionalrh,solaromanenum]{spr-sola-addons} % For Solar Physics 
\usepackage{graphicx}                    % For eps figures, newer & more powerfull
\usepackage{amssymb}                    % useful mathematical symbols
\usepackage{color}                       % For color text: \color command
\begin{document}

\begin{article}

\begin{opening}

\title{Temporal and Latitudinal Variation in Penumbra-Umbra Ratios of the Sunspots: Analyses of RGO, Kodaikanal and Debrecen Databases}

%%%%%%%%%%%%%%%%%%%%%%%%%%%%%%%%%%%%%%%%%%%%%%%%%%%
%% Authors Names
%
\author[addressref={aff1},corref,email={jouni.j.takalo@oulu.fi; jojuta@gmail.com}]{\inits{J.J.}\fnm{Jouni}~\lnm{Takalo}}

\institute{$^{1}$ Space Physics and Astronomy Research Unit, University of Oulu,
POB 3000, FIN-90014, Oulu, Finland\\
\email{jouni.j.takalo@oulu.fi}}

%\address[id=aff1]{ReSoLVE Centre of Excellence, Space Climate research unit, University of Oulu,
%POB 3000, FIN-90014, Oulu, Finland}
%\date{Received: }
%%%%%%%%%%%%%%%%%%%%%%%%%%%%%%%%%%%%%%%%%%%%%%%%%%%
%% Runningheads
%
\runningauthor{J.J. Takalo}
\runningtitle{Penumbra-Umbra Ratios of the Sunspots}

%%%%%%%%%%%%%%%%%%%%%%%%%%%%%%%%%%%%%%%%%%%%%%%%%%%
%% Affilations 
%% id shold be the same with \author addressref value.
%\address[id={}]{}

%%%%%%%%%%%%%%%%%%%%%%%%%%%%%%%%%%%%%%%%%%%%%%%%%%%
%%% Abstract 
\begin{abstract}
We study the latitudinal distribution and temporal evolution of the sunspot penumbra-umbra ratio ($q$)
for the even and odd Solar Cycles 12\,--\,24 of RGO sunspot groups, SC21\,--\,SC24 of Debrecen sunspot groups 
and Kodaikanal sunspot dataset for SC16\,--\,SC24.  We find that RGO even (odd) Cycles have $q$-values 5.20 (4.75), Kodaikanal even (odd) cycles have $q$-values 5.27 (5.43), and Debrecen cycles has $q$-value 5.74 on the average.  

We also show that $q$ is at lowest around the Equator of the Sun and increases towards higher latitudes having maximum values at about 10-25 degrees. This is understandable, because smaller sunspots and groups locate nearer to Equator and have smaller $q$-values than larger sunspots and groups, which maximize at about 10-20 degrees at both hemispheres. The error limits are very wide and thus the confidence of this result is somewhat vague.
 
For Debrecen dataset we find a deep valley in the temporal $q$-values before the middle of the cycle. We show that
this exists simultaneously with the Gnevyshev gap (GG) in the graph of the total and umbral areas of the large sunspot groups. Other databases do not show GG in their $q$-graphs, although GG exists in their temporal total area and umbral area.
\end{abstract}

%%%%%%%%%%%%%%%%%%%%%%%%%%%%%%%%%%%%%%%%%%%%%%%%%%%
%% Keywords
%
\keywords{Sun: Sunspot cycles, Sun: Sunspot area, Sun: Penumbra-umbra ratio, Method: Distribution analysis, 
Method: Statistical analysis}

\end{opening}
%-------------------------------------------------

%%%%%%%%%%%%%%%%%%%%%%%%%%%%%%%%%%%%%%%%%%%%%%%%%%%
%% Sections

\section{Introduction}

It has been known almost for two hundred years that the occurrence of sunspots is cyclic, although not strictly periodic. The length of the sunspot cycle (SC) has varied between 9.0 and 13.7 years. The shape of the sunspot cycle has also changed somewhat. Waldmeier noticed the asymmetry of sunspot cycles, with the ascending phase being typically shorter than the declining phase, and that there is an anti-correlation between cycle amplitude and the length of the ascending phase of the cycle \citep{Waldmeier_1935, Waldmeier_1939}. 
The number of sunspots has been observed, at least, since the 17th century, although the measurements in the early days were scarce and somewhat inaccurate. These early observations include also hand-drawn figures, but these are only for short periods \citep{Neuhauser_2018, Carrasco_2021, Carrasco_2022}. The positions and areas of the sunspots have been systematically recorded by the Royal Greenwich Observatory (RGO) since 1874 \citep{Hathaway_2013, Mandal_2020}. Recently, efforts have been done to estimate the sunspot areas and sometimes even trying to separate umbra and penumbra structures from the early drawings (see the excellent review by Artl and Vaquero, 2020). Since the late 19th century photographic recordings have been done in many observatories. As stated earlier the Royal Greenwich Observatory (RGO) started recordings in 1874 and continued these until 1976 \citep{Hathaway_2013}, Kodaikanal observatory in India has taken white-light images of the sunspots since 1904 \citep{Ravindra_2013, Mandal_2017b, Jha_2022}, Mt. Wilson has collections from 1917 up to 1982 \citep{Howard_1984,Bogdan_1988}, and Debrecen continued the recordings of RGO from 1974 until today \citep{Baranyi_2001, Gyori_2017}.

There have been some studies of penumbra-umbra ratio ($q$) of the sunspots during the 20th century. \cite{Nicholson_1933} and \cite{Waldmeier_1939} found consistent results with each other that $q$ decreased as the sunspot size increased. \cite{Jensen_1955} reported also a slight decrease of $q$-values as a function of sunspot size but that $q$ was higher during maxima than minima of the cycles. During cycle minima the variation of the $q$-values was also smaller with increasing sunspot size. \cite{Antalova_1971} noticed that $q$ is an increasing function of the sunspot size and in double maxima cycles the ratio is higher in the first maximum than in the second maximum. She also stated that $q$-values do not change with heliographic latitude.

\cite{Hathaway_2013} studied the daily records of sunspot group areas compiled by the Royal Observatory, Greenwich from May of 1874 through 1976. He found that, on average, $q$ increases from about 5 to 6 with the group area increasing from 100 to 2000$\,\mu$Hem, and does not change with latitude or phase of the cycle. He, however, found a peculiar change of $q$ with time such that it decreased smoothly from more than 7 in 1905 to lower than 3 by 1930 and then increased again to over 7 in 1961. \cite{Carrasco_2018} studied sunspot data from the catalogue published by the Coimbra Astronomical Observatory (CAO) for the period 1929\,--\,1941. They could not find such kind of change during the analyzed period. There were also significant differences in $q$-values for the smaller groups between the RGO and CAO analyses. They state that the main differences are in the measurements of the umbra area, and while CAO instrument and methods did not change during the period 1929\,--\,1941, there were changes in RGO arrangements, which could have affected the discrepancy of the results.

\cite{Jha_2019} and \cite{Jha_2022} have studied $q$-values from the recently published Kodaikanal digitized white-light images from 1904 to 2017. They have also compared their results to the $q$-values of RGO and Debrecen datasets for the same period. They found that $q$ increases from 5.5 to 6 as the sunspot size increases from 100 to 2000\,$\mu$Hem. They did not find any systematic trend in smaller ($<$100\,$\mu$Hem) sunspots as was reported earlier in the results from RGO database by \cite{Hathaway_2013}. Furthermore, they found that the average $q$ does not show variation as a function of latitude. They compare $q$ time-series calculated from Kodaikanal dataset to the $q$ time-series calculated from RGO and Debrecen sunspot datasets. It seems that for the sunspots $<$100\,$\mu$ Hem, $q$-values for Kodaikanal data fluctuate between 4 to 5 all the time and for sunspots $\geq$100\,$\mu$Hem between 5 to 6, while for the other two datasets (RGO and Debrecen) the variation is much larger. So Kodaikanal sunspot dataset seems to be more homogeneous than the other two datasets.

Recently \cite{Hou_2022} published an article where they analyze the recently digitized sunspot drawings observed from Yunnan Observatories (1957\,--\,2021) in order to study $q$ for the Solar Cycles 19\,--\,24. They found that $q$ is 6.63 $\pm$ 0.98, and its probability distribution fits very well to the log-normal distribution function. They did not notice any dependence of the $q$-value on the latitude or the phase of the cycle.
 
In this study we analyze the latitudinal distribution and temporal evolution of sunspot penumbra-umbra ratio ($q$) of SC12\,--\,SC23 (occasionally also SC24, which is, however, only until 2017). To get more data for solid statistics, we superimpose separately the even and odd cycles. To this end we harmonize the duration of the cycles such that all have the same length 128 months, which is about the average for the Solar Cycles 16\,--\,23, i.e. the interval for the Kodaikanal dataset. This paper is organized as follows. Section 2 presents the databases used in this study. In Section 3 we study the $q$-values of Debrecen dataset for Cycles 21\,--\,24. Section 4 deals with the $q$-values of RGO Cycles 12\,--\,20 and Section 5 the $q$-values of Kodaikanal Cycles 16\,--\,24. In Section 6 we compare the results for the overlapping intervals of the aforementioned datasets and give our conclusions in Section 7.

\section{Sunspot Area Databases}

In the analyses of sunspot area we use three databases. The longest dataset for umbral and total sunspot areas is recently published Kodaikanal database \citep{Mandal_2017b, Jha_2022}, which is reconstructed from the white-light images of the same observatory since 1904. Because there are gaps before 1921, and Cycle 24 is incomplete, we use this dataset mainly for the Solar Cycles 16\,--\,23. Notice that this data contains areas of separate sunspots. Another database is that of the Royal Observatory, Greenwich-USAF/NOAA Sunspot Data for the years 1874\,--\,2016. This database contains, among others, time, latitude and whole area size (in millionths of solar hemisphere, $\mu$Hem) for individual sunspots for SC12\,--\,SC23, and also umbra sizes of the sunspots for SC12\,--\,SC20 \citep{Hathaway_2013}. We use here the data for Solar Cycles 12\,--\,20, i.e the period which contains the whole and umbra areas of the sunspot group data. The third dataset is the Debrecen Photoheliographic Data (DPD), which consist of daily, group and sunspot data \citep{Baranyi_2001, Gyori_2017}. Here we use whole areas and umbra areas of the sunspot group data data, marked in the database with ``g". Because Debrecen database seems to be most precise for the recent Solar Cycles 21\,--\,24, we use it as a preliminary database. The minima and lengths for the Solar Cycles 12\,--\,24 used in this study are listed in Table 1.

\begin{table}
\small
\caption{Sunspot-cycle lengths and dates [fractional years, and year and month] of (starting) sunspot minima for Solar Cycles 12\,--\,24 \citep{NGDC_2013}.}
\begin{tabular}{ c  c  l  c }
  Sunspot cycle    &Fractional    &Year and month     &Cycle length  \\
      number    &year of minimum   & of minimum     &    [years] \\
\hline
12    & 1879.0  &1878 December & 10.6 \\				
13    & 1889.6  &1889 August & 12.1 \\			
14    & 1901.7  &1901 September  & 11.8  \\
15    & 1913.5  &1913 July  &  10.1  \\
16    & 1923.6  &1923 August  & 10.1  \\			
17    & 1933.7  &1933 September & 10.4 \\			
18    & 1944.1  &1944 February  & 10.2  \\
19    & 1954.3  &1954 April  & 10.5  \\
20    & 1964.8  &1964 October  & 11.7  \\
21    & 1976.5  &1976 June & 10.2  \\
22    & 1986.7  &1986 September  & 10.1  \\
23    & 1996.8  &1996 October  & 12.2  \\
24    & 2009.0  &2008 December  & 10.9  \\ 
25    & 2020    &2019 December 

\end{tabular}
%\label{cycles}
\end{table}

In this study we define the penumbra-umbra ratio ($q$) as \citep{Antalova_1971, Hathaway_2013}
\begin{equation}
	q = \left(A_{\rm Sp}-A_{\rm Um}\right)/A_{\rm Um} ,
\end{equation}
where $A_{\rm Sp}$ is the sunspot total area and $A_{\rm Um}$ the umbral area of the sunspot.

\section{Debrecen Sunspot Group Dataset}

\begin{figure}
	\centering
	\includegraphics[width=1.0\textwidth]{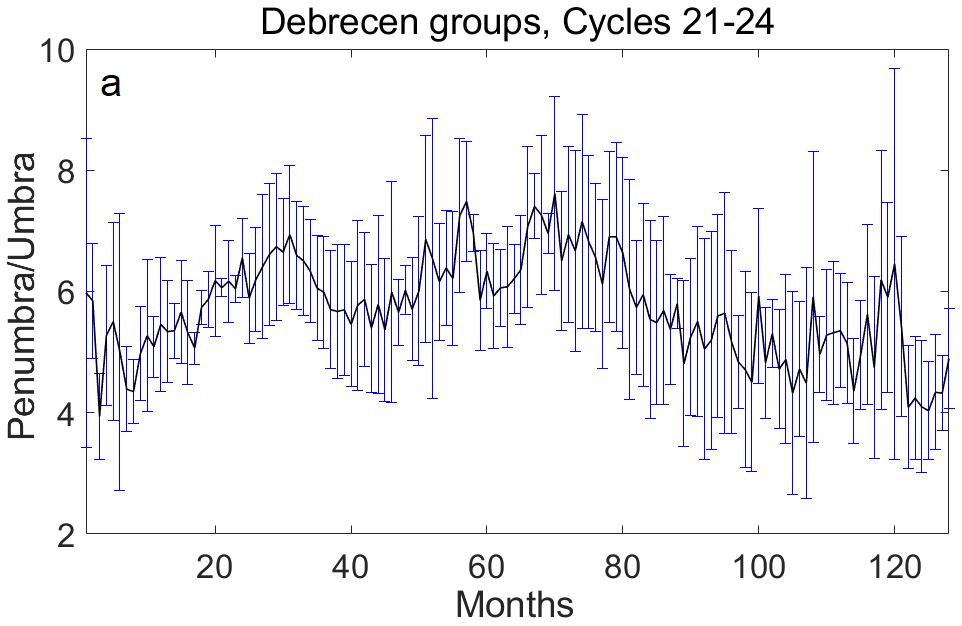}
		\caption{Debrecen temporal penumbra-umbra ratio of all sunspots for Cycles SC21\,--\,24 (black curve). The blue errorbars show the standard deviations at each monthly point of the ratio.}
		\label{fig:Debrecen_penumbra_umbra_temporal}
\end{figure}

\begin{figure}
	\centering
	\includegraphics[width=1.0\textwidth]{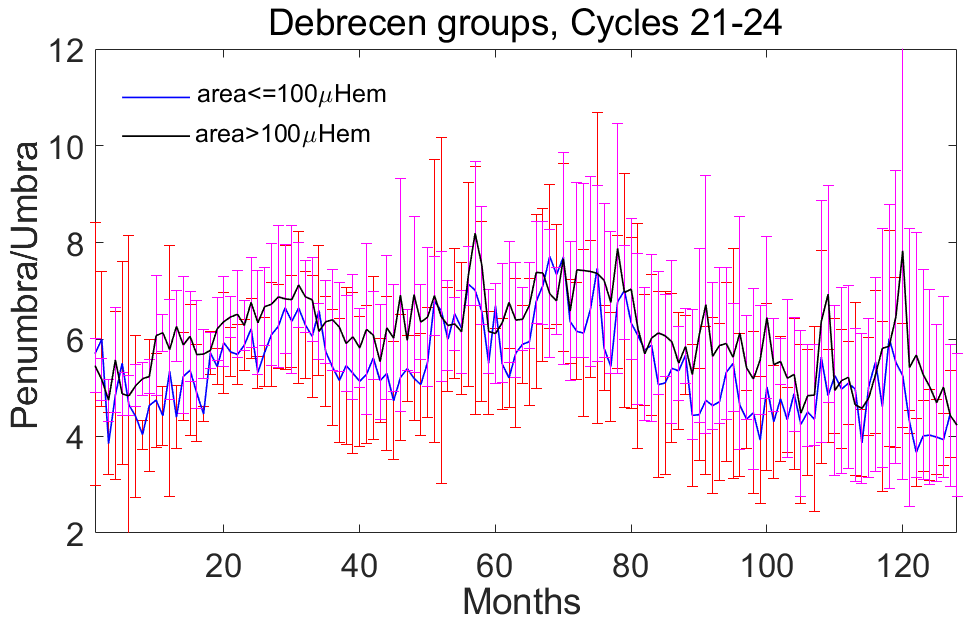}
		\caption{Debrecen temporal penumbra-umbra ratio of the sunspot groups for Cycles SC21\,--\,24  separately for areas $\leq$100\,$\mu$Hem and $>$100\,$\mu$Hem as blue and black curves, respectively. The errorbars are shown with red and magenta colors, respectively.}
		\label{fig:Debrecen_small_large_area ratios}
\end{figure}

\begin{figure}
	\centering
	\includegraphics[width=1.0\textwidth]{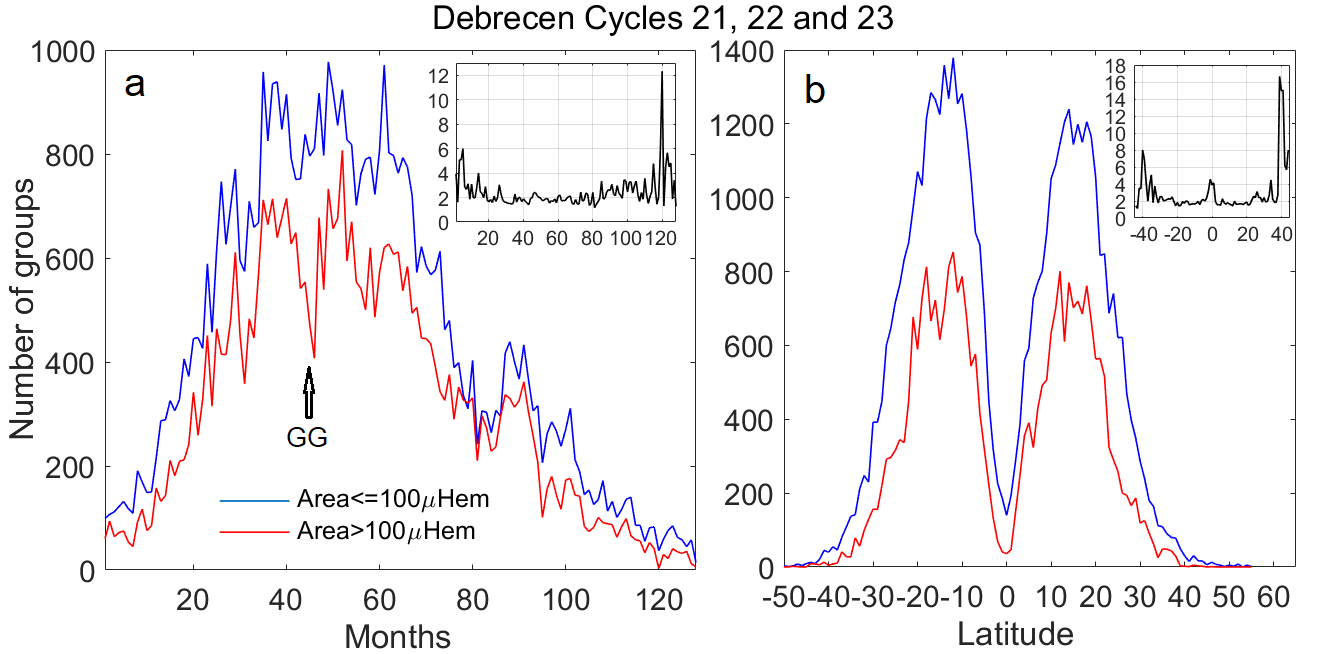}
		\caption{a) Number of small (blue) and large (red) sunspot groups as a function of time for Debrecen Solar Cycles 21\,--\,23. The inset shows the ratio of small/large groups along the cycles. b) Number of small (blue) and large (red) sunspot groups as a function of latitude for Debrecen Solar Cycles 21\,--\,23. The inset shows the ratio of small/large groups along the latitudes.}
		\label{fig:Debrecen_number_of_groups}
\end{figure}

Figure \ref{fig:Debrecen_penumbra_umbra_temporal} shows the temporal distribution of penumbra-umbra ratio, q, for Solar Cycles 21\,--\,24 (Cycle 24 only until 19th June, 2018) of Debrecen sunspot group data. In order to get more data to a more solid statistics, we harmonize the solar cycles having each 128 months, which is about their average cycle-length during the longest database in this study, i.e. Kodaikanal Solar Cycles 16\,--\,24. We then study separately the superimposed cycles of the aforementioned period for Debrecen data data. The blue vertical errorbars show the standard deviation at each point of the monthly measurements. Interestingly there is a huge valley at the maximum region of the cycles, i.e. around 40-50 months. To analyze this valley in more detail we plot separately the temporal distributions for areas smaller or equal to 100\,$\mu$Hem (small) and over 100\,$\mu$Hem (large). Figure \ref{fig:Debrecen_small_large_area ratios} shows $q$ for small and large category sunspot groups as blue (red errorbars) and black (magenta errorbars) curves, respectively. The valley exists in both categories but are somewhat deeper in the small category groups (when studying in more detail, it turned out that the decrease is deepest for groups between 50\,--\,100\,$\mu$Hem). We believe that this decrease in the $q$ is related to the Gnevyshev gap (GG) \citep{Gnevyshev_1967} at about 35\,--\,40\,\% from the start of the cycles \citep{Takalo_2018}. Figure \ref{fig:Debrecen_number_of_groups}a shows the number of small (blue) and large (red) sunspot groups during Cycles 21\,--\,23 (Cycle 24 is omitted, because it is incomplete). Note that the number of large groups decreases more than the number of small groups during the GG. This result is consistent with the result by \cite{Takalo_2020_1} that GG is more visible in large than in small sunspots. It is then understandable that when the relative number of smaller sunspot groups increases, the $q$-values are lower at the GG-region. There is another drop after 60 months but it is caused mainly by Cycle 24, when large groups disappear abruptly after the cycle maximum on 2014. Although the errorbars are quite wide, the shape of the penumbra-umbra curves are very obvious regarding that small groups dominate at the beginning and at the end of the cycles and large groups are relatively more abundant during the maximum of the cycles (see the inset of Figure \ref{fig:Debrecen_number_of_groups}a). The mean $q$-value for small groups ($\leq$100\,$\mu$Hem) is 5.44 (standard deviation, std=0.90), for large groups ($>$100\,$\mu$Hem) 6.09 (std=0.82) and for all groups (Fig. \ref{fig:Debrecen_penumbra_umbra_temporal}) 5.74 (std=0.82). Note that the standard deviation in the brackets here and later is different from the errorbars in the figures, i.e. it is calculated from the average monthly (or average latitudinal) $q$-values during the cycles, i.e. the larger this std is the more variation (trends) there is in the distribution in the $q$-values.

\begin{figure}
	\centering
	\includegraphics[width=1.0\textwidth]{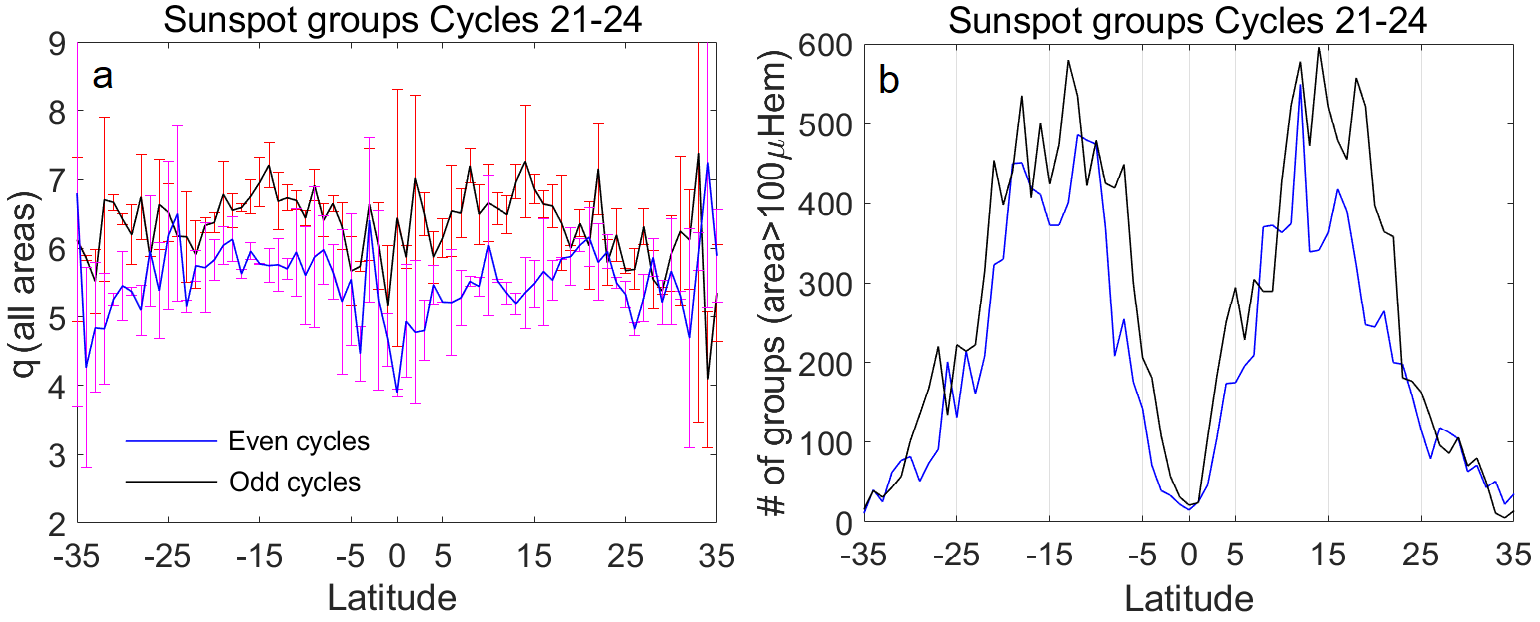}
		\caption{a) Debrecen latitudinal $q$ for all sunspots of the even (blue with magenta errorbars) and odd (black with red errorbars) Cycles 21\,--\,24. b) Latitudinal distribution for the number of sunspot groups of the even (blue) and odd (black) Cycles 21\,--\,24.}
		\label{fig:Debrecen_penumbra_umbra_latitudinal}
\end{figure}

Figure \ref{fig:Debrecen_penumbra_umbra_latitudinal}a shows the Debrecen latitudinal distribution for the even and odd cycles SC21\,--\,SC24. The $q$-graph for the even cycles has a valley around zero latitude, except that of different values at -3 degrees of latitude. The maximum values seem to be between 10 to 20 degrees at both hemispheres. This is also the region, where the larger sunspot ($>$100\,$\mu$Hem) are most abundant (see Figure \ref{fig:Debrecen_penumbra_umbra_latitudinal}b). It is, however, clear that the penumbra-umbra ratio is smaller near the Equator of the Sun. This is due to the smaller size of sunspots dominating near the Equator, which leads also to a smaller $q$-value. Note from the inset of Figure \ref{fig:Debrecen_number_of_groups}b that the ratio of small/large sunspots is about four between -2\,--\,2 degrees while is is under 2 from 5 to 25 degrees in both hemispheres. It is also evident that the odd cycles have higher $q$-values than the even cycles. This is because the odd cycles (21, 23) have more larger category sunspot groups than the even cycles (22, 24). Figure \ref{fig:Debrecen_penumbra_umbra_latitudinal}b shows the latitudinal number of large sunspot group for even (blue) and odd (black) cycles. Note that the difference of $q$-values is largest at the latitudes in which the difference in number of groups is also largest. Figure \ref{fig:Debrecen_penumbra_umbra_latitudinal}b shows also that the number of large groups has a local minimum at or near 15 degrees of latitude. This is especially clear for even cycles. This is consistent with the results of \cite{Takalo_2020_1, Takalo_2020_2}, that large sunspots and sunspot groups have local minima at the time when the average latitude of sunspots crosses 15 degrees of latitude \citep{Takalo_2020_2}. We believe that this in turn coincides with the Gnevyshev gap in the SSN and sunspot areas. The mean $q$-values for the sunspot groups of even and odd Cycles 21\,--\,24 between latitudes -35\,--\,35 are 5.53 (std=0.56) and 6.33 (std=0.55), respectively.

\section{Kodaikanal Sunspot Dataset}

\begin{figure}
	\centering
	\includegraphics[width=1.0\textwidth]{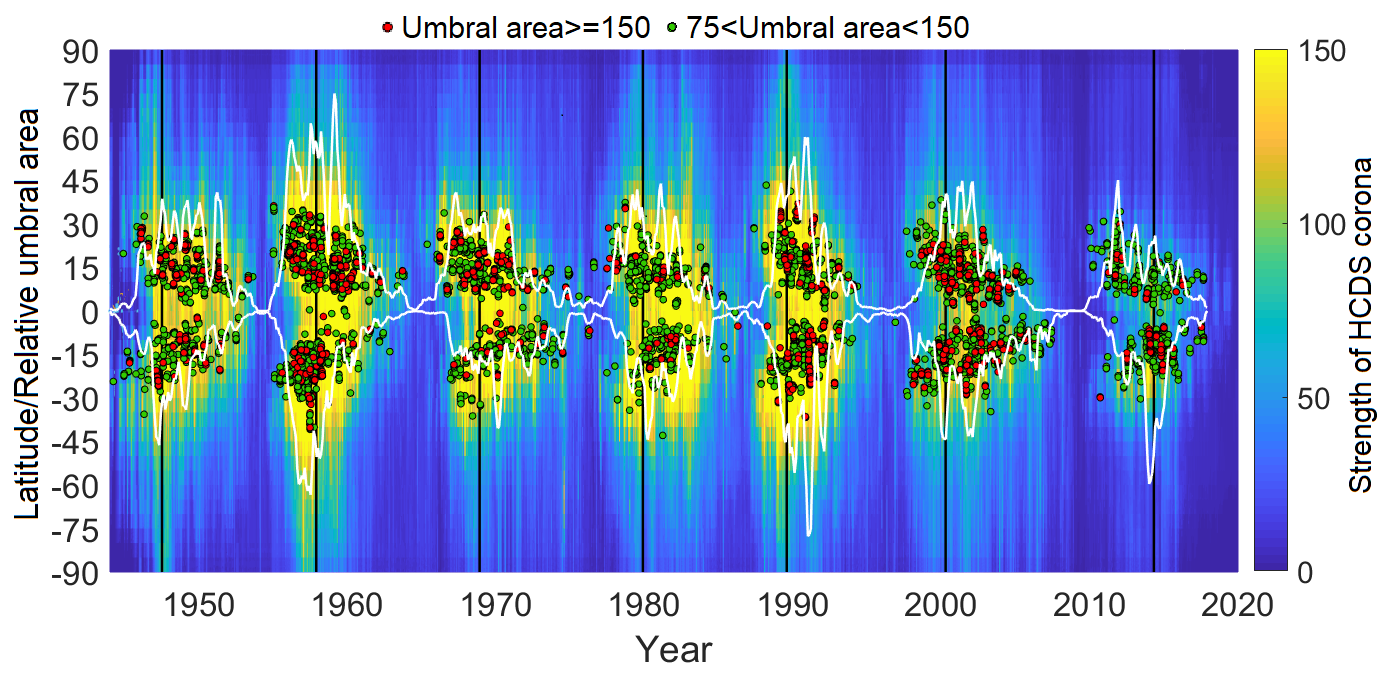}
		\caption{The butterfly diagram of Kodaikanal dataset umbral areas for Solar Cycles 18\,--\,24. The green and red dots are sunspots with umbral area 75 to 150$\,\mu$Hem and greater or equal than 150$\,\mu$Hem, respectively. The slightly smoothed white curves show total umbral area for northern and southern hemisphere of the Sun. The colors on the base of the figure show HCDS corona, whose intensities are shown on the right. The black vertical lines show the sites of the maxima of the cycles.}
		\label{fig:UmbrasAndHCDS}
\end{figure}

We start the study of the Kodaikanal dataset with a plot of the hemispheric umbral areas in Figure \ref{fig:UmbrasAndHCDS} for Solar Cycles 18\,--\,24 (note that Cycle 24 is incomplete). The reason why we concentrate in this figure on this period is the base-map, which shows the so-called Homogeneous Coronal Data Set (HCDS). The HCDS is the irradiance of the Sun as a star in the coronal green line (Fe \textsc{XIV}, 530.3\,nm). It is derived from ground-based observations of the green corona made by the network of coronal stations (Kislovodsk, Lomnick\'{y} \v{S}t\'{i}t, Norikura, and Sacramento Peak). The coronal intensities have been measured at 72 points at five-degree separation starting from North Pole counterclockwise around the Sun at a height around 50 arcsec. The values are calibrated to the center of the solar disk to get absolute values of intensity, i.e. absolute coronal units (ACU). One ACU represents the intensity of the continuous spectrum of the center of the solar disk in the width of one {\AA}ngstr{\"o}m at the same wavelength as the observed coronal spectral line (1\,ACU = 3.89\,Wm$^{-2}$ sr$^{-1}$ at 530.3\,nm) \citep{Minarovjech_2011, Takalo_2022_2}. This database exists only for SC18\,--\,24 and the temporal and spatial intensities of this corona are shown as a colorbar on the right-side of the figure. The green and red dots are sunspots with umbral area 75 to 150$\,\mu$Hem and greater or equal than 150$\,\mu$Hem, respectively (we do not show the smaller sunspots here for clarity such that the corona is also visible). The white curves show the (relative) hemispheric total umbral areas, and the vertical black lines the official maxima of the solar cycles. Significant information can be extracted from Figure \ref{fig:UmbrasAndHCDS}. It is evident that the corona and stronger sunspots are located simultaneously, although corona covers larger latitudinal region. It is seen that the maximal corona starts at latitudes 35\,--\,40 degrees similarly to sunspots, but migrates both towards the Equator and poleward during cycle evolution, although as fainter toward the Poles than equatorside. It is also seen that the strongest Solar Cycle 19 is most symmetric around the cycle in corona and sunspot in the northern hemisphere, while most of the largest sunspots in the southern hemisphere are located in the ascending phase of the cycle. The other cycles, except Cycle 24, have a prolonged tail such that the majority of the large sunspots are after the cycle maximum. Solar Cycle 24 has hemispheric asymmetry such that the largest northern hemisphere sunspots exist before the cycle maximum and southern hemisphere sunspots locate around the cycle maximum. Note also, that Solar Cycles 23 and 24 have much fainter HCDS corona than other cycles.

\begin{figure}
	\centering
	\includegraphics[width=1.0\textwidth]{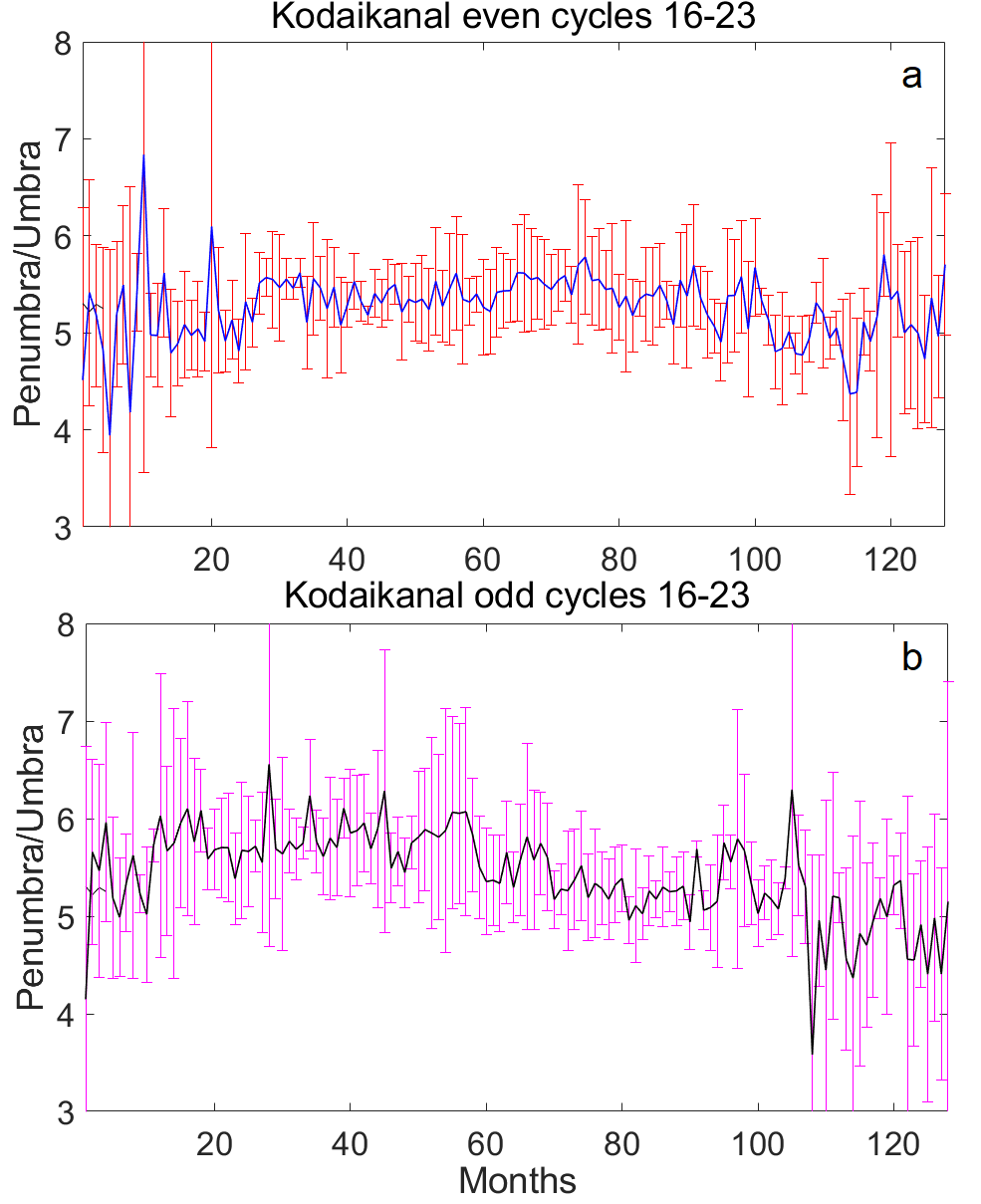}
		\caption{a) Kodaikanal temporal penumbra-umbra ratio ($q$) for all sunspots of even Cycles 16\,--\,23. b) $q$ for all sunspots of odd Cycles 16\,--\,23.}
		\label{fig:Koda_penumbra_umbra_temporal}
\end{figure}

\begin{figure}
	\centering
	\includegraphics[width=1.0\textwidth]{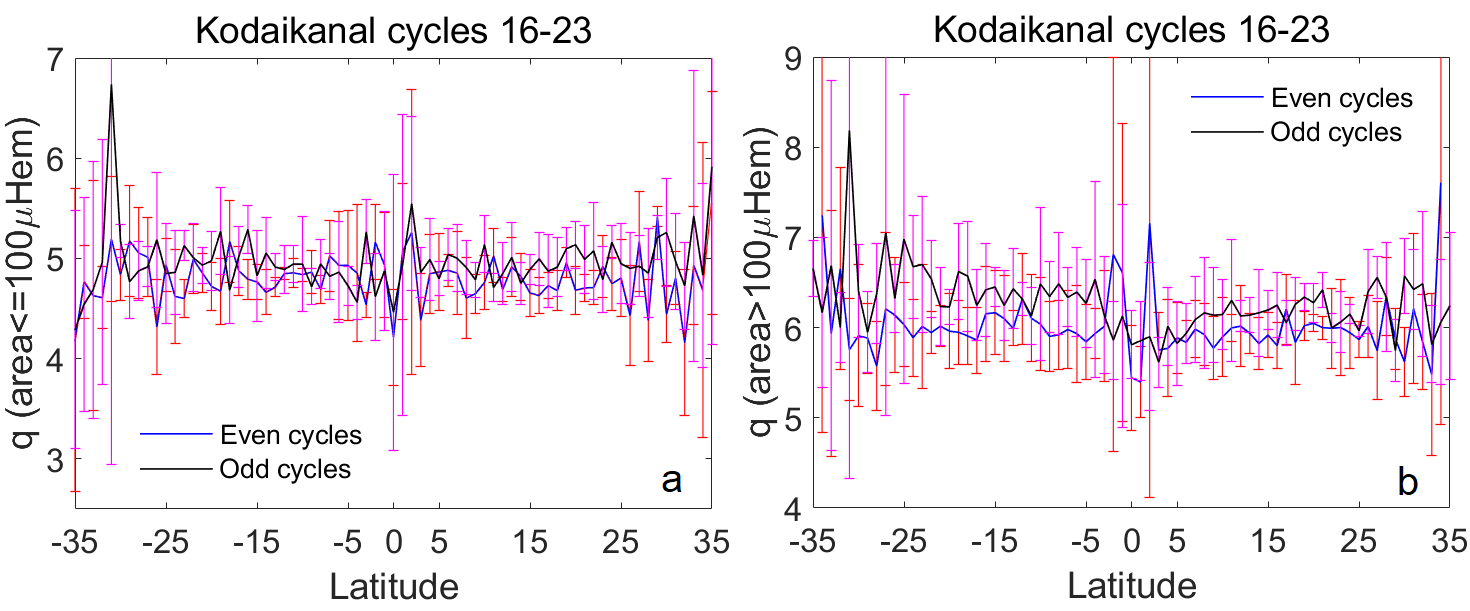}
		\caption{a) Kodaikanal latitudinal penumbra-umbra ratio ($q$) of all sunspots smaller than 100\,$\mu$Hem  for even and odd Cycles 16\,--\,23. b) Latitudinal $q$ for sunspots larger than 100\,$\mu$Hem for even and odd Cycles 16\,--\,23.}
		\label{fig:Koda_penumbra_umbra_latitudinal}
\end{figure}

The recently published Kodaikanal sunspot dataset is the longest sunspot set containing also areas of sunspot umbras recorded on a single station. We again harmonize Solar Cycles 16\,--\,23 having each 128 months. Figures \ref{fig:Koda_penumbra_umbra_temporal}a and b show the penumbra umbra ratio ($q$) for the even and odd cycles for all sunspots of SC16\,--\,SC23. These figures show that the odd cycles have moderately higher ratio than the even cycles in the ascending phase of the average cycle. This means that odd cycles have considerable more large sunspots in the ascending phase than after the maximum. On the other hand the shape of the $q$-graph for the even cycles reminds more the shape of the sunspot cycle itself, i.e, largest sunspots are in the middle of the cycle. The error limits are smaller than for Debrecen data, especially for the even cycles. The mean $q$ for even and odd cycles are 5.27 and 5.43 with standard deviations 0.36 and 0.47 calculated from the monthly $q$-values, respectively.

Figures \ref{fig:Koda_penumbra_umbra_latitudinal}a and b show the latitudinal penumbra-umbra ratio for sunspots smaller than 100\,$\mu$Hem and for sunspots larger than 100\,$\mu$Hem of even (blue with red errorbars) and odd (black with magenta errorbars) cycles. We restrict again the latitudes between -35 and 35 degrees, because the small number of umbral data (the divisor in the calculation of penumbra-umbra ratio) causes quite wild behavior at larger latitudes. Note that the error limits are nevertheless large around zero latitude and at latitudes higher than 25 degrees. The latitudinal distribution of $q$-values is quite flat for small sunspots but have more variation for large sunspots. Note that for large category odd cycles there is again a valley around zero latitude, but there are bad values (very wide errorbars) on both sides of the zero latitude for the even cycles. The mean values for $q$ between -35 to 35 degrees of smaller category and larger category sunspots are for the even cycles 4.79 (std=0.32) and 6.02 (0.37), and for the odd cycles 5.00 (std=0.48), 6.29 (0.36), respectively.

\section{RGO Sunspot Group Dataset}

\begin{figure}
	\centering
	\includegraphics[width=1.0\textwidth]{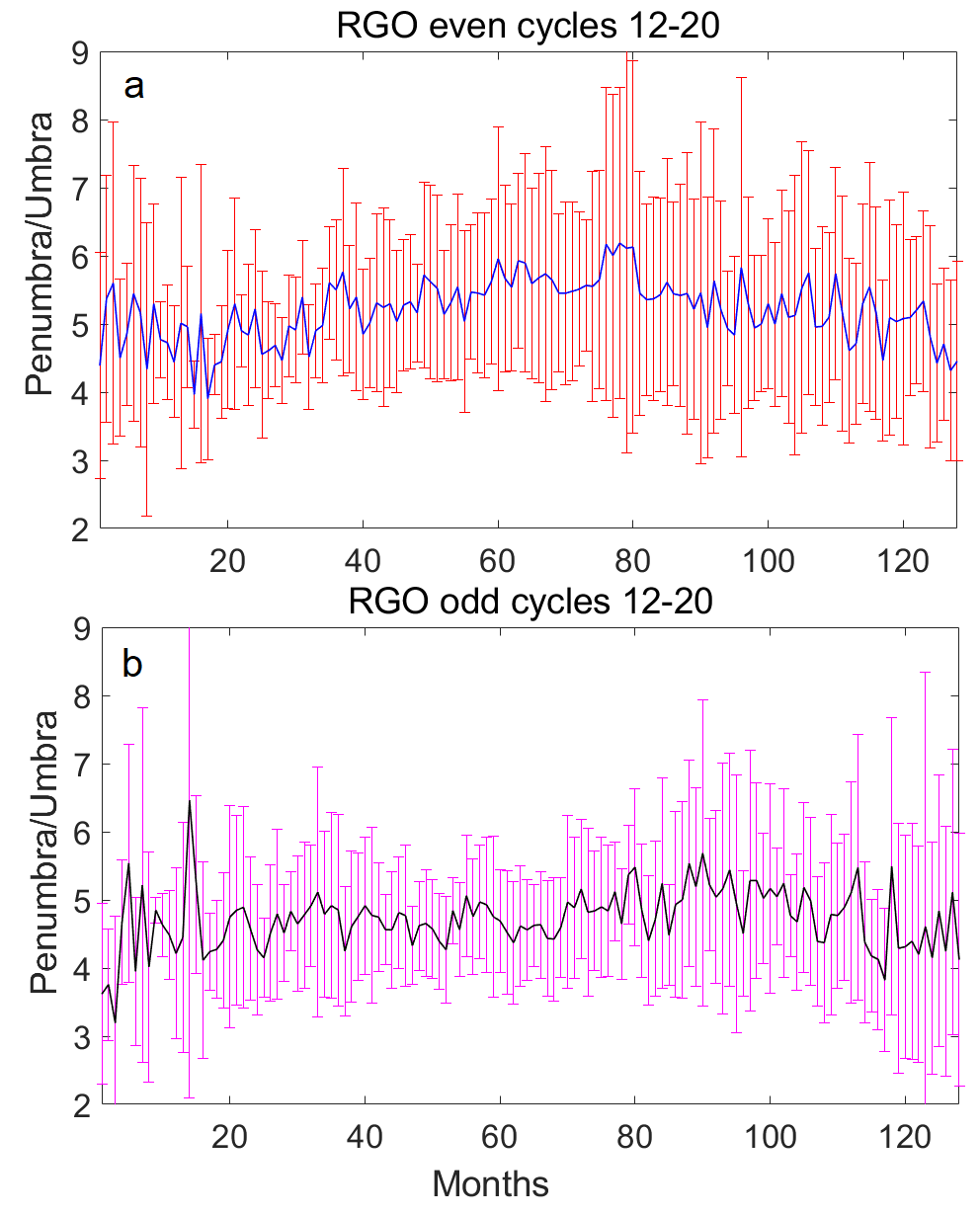}
		\caption{a) RGO temporal penumbra-umbra ratio ($q$) of all sunspots for the even Cycles 12\,--\,20. b) RGO temporal $q$ of all sunspots for the odd Cycles 12\,--\,20.}
		\label{fig:RGO_penumbra_umbra_temporal}
\end{figure}

The Royal Greenwich Observatory (RGO) compiled sunspot group observations from a small network of observatories starting in May of 1874. These observations lasted until 1976 after which the measurement have been done in Debrecen. The RGO sunspot group dataset contains also umbra areas and that is why we can study penumbra-umbra ratios for the total Solar Cycles 12\,--\,20 using this dataset. We use the same common length for the cycles in order to compare the results with Kodaikanal distributions. Figures \ref{fig:RGO_penumbra_umbra_temporal}a and b  show the temporal penumbra-umbra ratios of the RGO dataset for the even and odd cycles, respectively. In this case $q$ for the even cycles seems to be somewhat higher than for the odd cycles. The monthly $q$-graph is flatter for the odd cycles than for the even cycles, but both have slightly higher values in the descending part of the cycles. The mean values for $q$ are 5.20 and 4.75 with standard deviations 0.46 and 0.41 for the even and odd cycles, respectively. Note that now (Cycles 12\,--\,20) even cycles have higher average $q$-value than odd cycles, while for Kodaikanal (Cycles 16\,--\,23) even cycles have lower average $q$-value than odd cycles.

\begin{figure}
	\centering
	\includegraphics[width=1.0\textwidth]{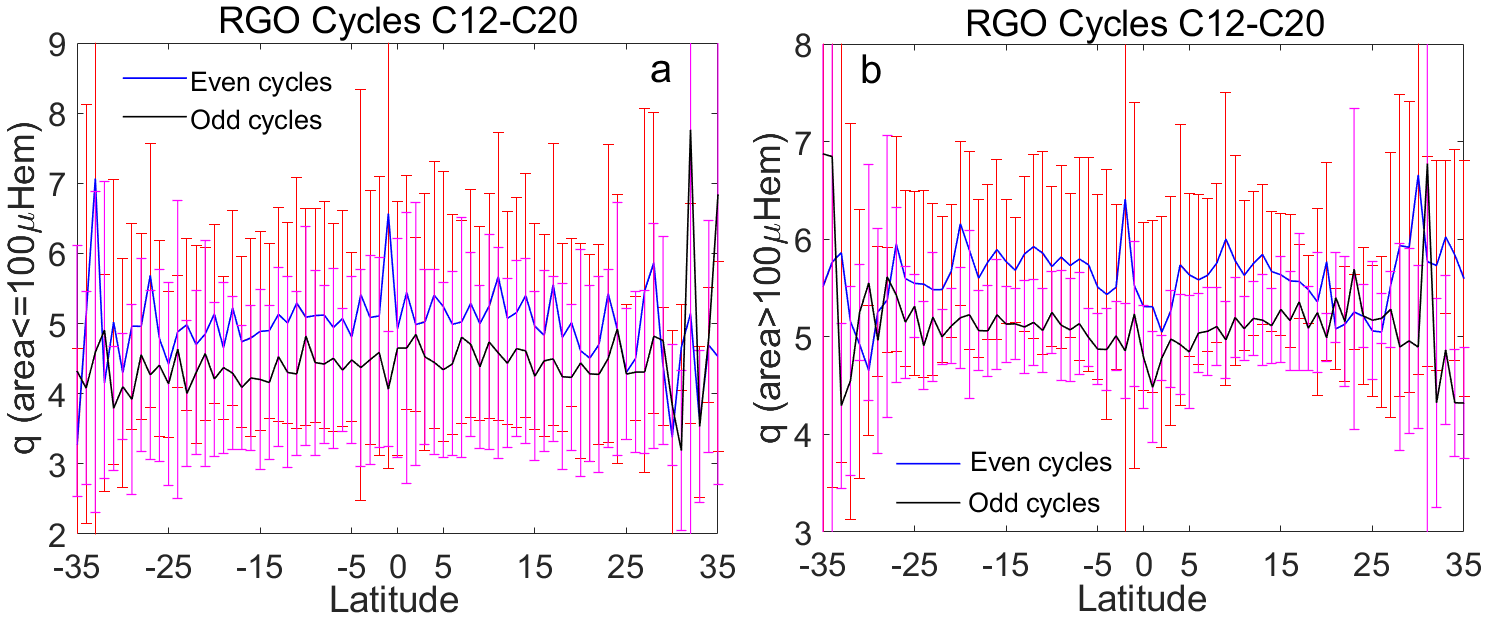}
		\caption {a) RGO latitudinal penumbra-umbra ratio ($q$) of sunspots smaller than 100\,$\mu$Hem for even and odd Cycles 12\,--\,20. b) RGO latitudinal $q$ for sunspots larger than 100\,$\mu$Hem for even and odd Cycles 12\,--\,20.}
		\label{fig:RGO_penumbra_umbra_latitudinal}
\end{figure}

Figures \ref{fig:RGO_penumbra_umbra_latitudinal}a and b show the latitudinal penumbra-umbra ratios for sunspot groups smaller than 100\,$\mu$Hem and larger than 100\,$\mu$Hem, respectively. The overall shape of the ratio for small groups is flat, if not slightly convex upwards with no valleys or humps. There are, however, huge errorbars in the small category plots. These are due to the large variation in the RGO penumbra/umbra ratio, especially for small groups \citep{Hathaway_2013, Jha_2022}. As stated earlier, \cite{Carrasco_2018} did not find such large variation in the study of Coimbra Astronomical Observatory measurements for the period 1929\,--\,1941. For larger sunspot groups there is a shallow valley around zero degrees, except a couple of anomalous values at negative low latitudes. The maxima are between 5 to 20 degrees of latitude for the even cycles, but are wider and flatter for the odd cycles. Note that, although there are valleys around zero latitude, they have just minor significance, because the errorbars for large group $q$-values are also quite wide. The $q$-values between -35 to 35 degrees for small and large groups are 4.97 (std=0.56), 4.47 (0.58) and 5.61 (0.33), 5.13 (0.45) for the even and odd cycles, respectively.

\section{Comparison of Kodaikanal, RGO and Debrecen Sunspot Datasets}

\begin{figure}
	\centering
	\includegraphics[width=1.0\textwidth]{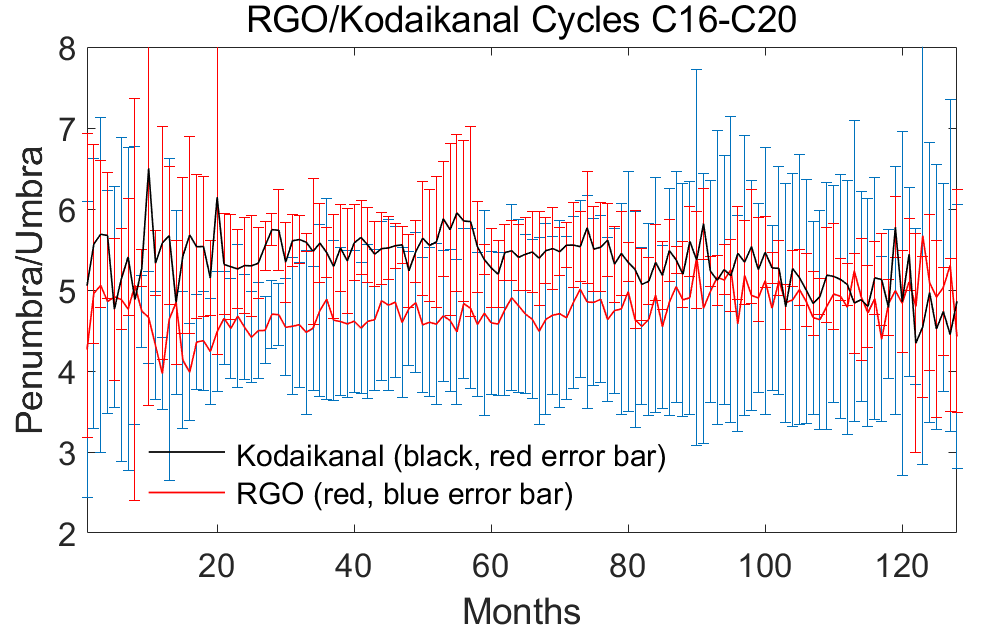}
		\caption{Comparison of Kodaikanal and RGO temporal penumbra-umbra ratios for Solar Cycles 16\,--\,20.}
		\label{fig:RGO_Koda_penumbra_umbra_comparison}
\end{figure}

There are overlapping Cycles 16\,--\,20 in RGO and Kodaikanal datasets and Cycles 21\,--\,23 in Kodaikanal and Debrecen dataset. That is why it is appropriate to compare these intervals between those data. Figure \ref{fig:RGO_Koda_penumbra_umbra_comparison} shows the penumbra-umbra ratios for RGO and Kodaikanal Solar Cycles 16\,--\,20. It is clear that Kodaikanal $q$ is moderately higher than RGO $q$, except and the end of the average cycle. The shape of the $q$-graph for Kodaikanal is quite similar to Kodaikanal even cycles (Figure \ref{fig:Koda_penumbra_umbra_temporal}) but the $q$-graph of the RGO has no trends. This is because the $q$-graphs for separate cycles have opposite trends (C16 decreasing, C17 increasing etc., see also Jha et al., 2022), which cancel each other when they are overlaid such that the sum is quite flat. The mean values for the average Cycle between 16\,--\,20 are 5.35 and 4.75 with standard deviations 0.33 and 0.26 for Kodaikanal and RGO cycles, respectively.

\begin{figure}
	\centering
	\includegraphics[width=1.0\textwidth]{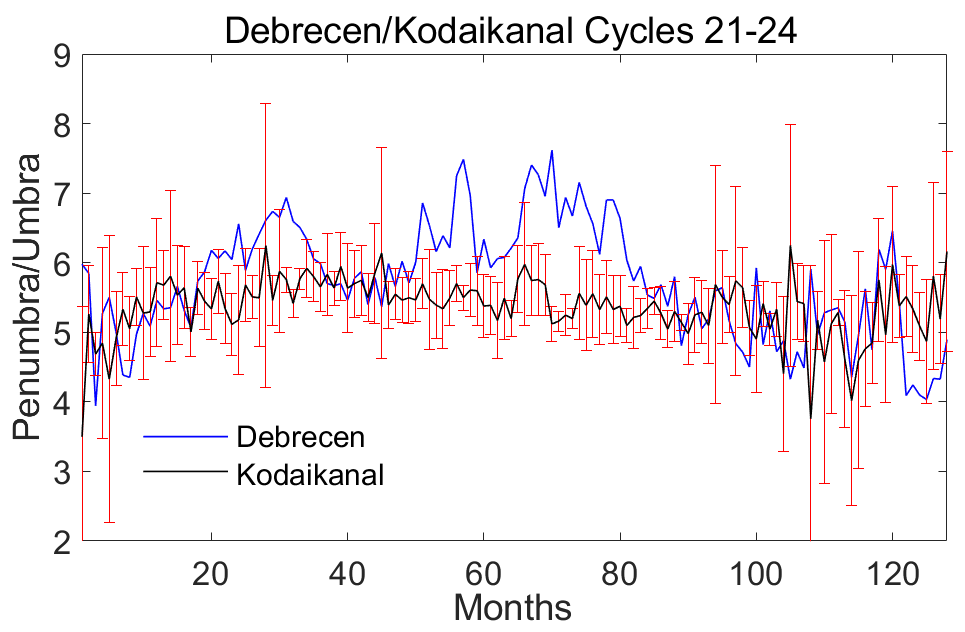}
		\caption{Comparison of Kodaikanal and Debrecen temporal penumbra-umbra ratios for Solar Cycles 21\,--\,24.}
		\label{fig:Koda_Deb_penumbra_umbra_comparison}
\end{figure}

Figure \ref{fig:Koda_Deb_penumbra_umbra_comparison} shows the $q$-graphs for Kodaikanal and Debrecen sunspot datasets of Solar Cycles 21\,--\,24. We show only the errorbars of Kodaikanal $q$-values, because the graph for Debrecen is same as in Figure \ref{fig:Debrecen_penumbra_umbra_temporal}. It is evident that Debrecen has higher values except at the very beginning and at the end of the cycles. Otherwise the $q$-graph of Debrecen data has much more structure than the $q$-graph of Kodaikanal data. The most interesting difference between these $q$-graphs is the GG-related decrease around 40-50 months in Debrecen $q$, which does not exist in Kodaikanal $q$. Note that Kodaikanal $q$ is highest in the ascending phase of the cycle and has also a shallow valley in $q$-values somewhat later than Debrecen, i.e. between 50\,--\,60 months. The mean $q$-value for Debrecen is 5.90 (std=0.76), and mean $q$-value for Kodaikanal is 5.34 (std=0.56).

\begin{figure}
	\centering
	\includegraphics[width=1.0\textwidth]{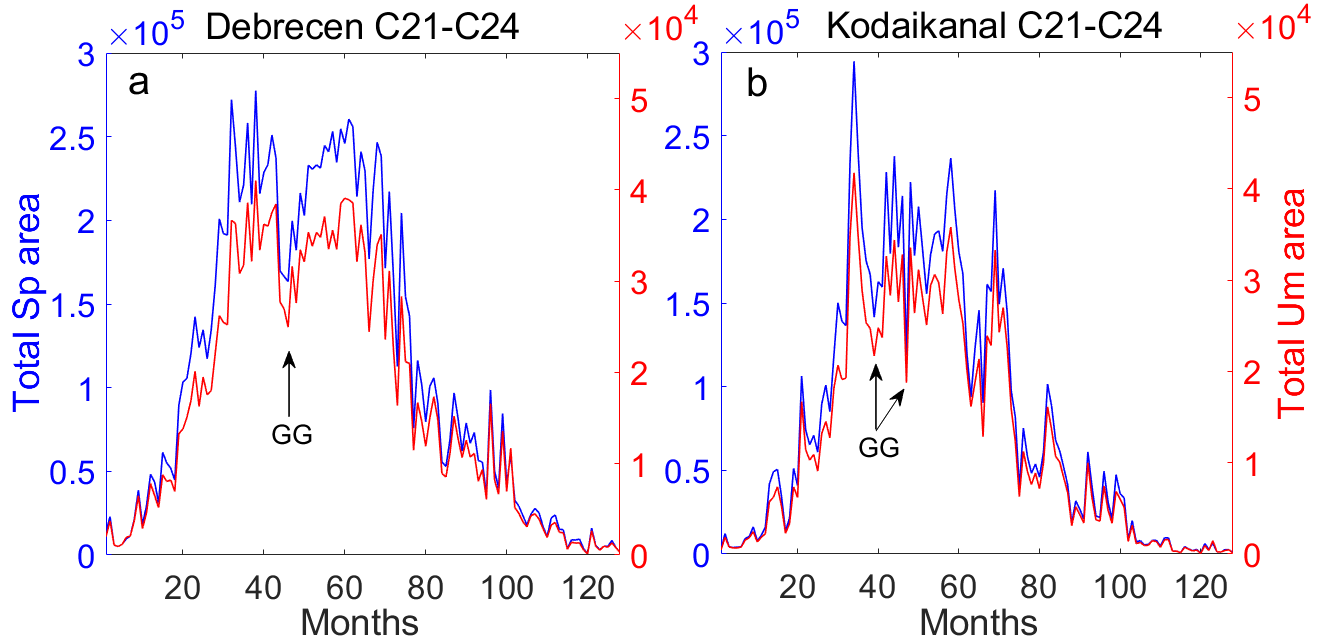}
		\caption{a) Total sunspot group areas and group umbral areas of Debrecen data for Solar Cycles 21\,--\,24. b) Total sunspot areas and sunspot umbral areas of Kodaikanal data for Solar Cycles 21\,--\,24. (Sunspot/group areas as blue color with axis on the left side and umbral areas as red with axis on the right side)}
		\label{fig:Koda_Deb_total_areas}
\end{figure}

Figures \ref{fig:Koda_Deb_total_areas}a and b show the total and umbral areas for Debrecen and Kodaikanal data, respectively. This figure may explain why Kodaikanal $q$ does not show the GG-related valley, which is so distinct in the Debrecen temporal $q$. Debrecen data have a deep decrease in both total sunspot and umbral area between 42\,--\,51 months, which is related to GG-phenomenon. On the other hand, Kodaikanal data have two drops, first at 35\,--\,42 and second very short at Debrecen GG region (47 months). It turns out that the first is related to odd cycles and the second to even cycles. The latter drop is narrow, because odd cycles have maxima at the time of the even cycles minima. Neither of these seem, however, to cause the $q$-value to decrease during the GG-regions. Note that Kodaikanal areas are calculated from separate sunspots and Debrecen areas from sunspot groups. We, however, have done similar a analysis also to Debrecen sunspot data (shown with ``s" in the database), and it gives results consistent with the Debrecen $q$-values of the groups.

\begin{figure}
	\centering
	\includegraphics[width=1.0\textwidth]{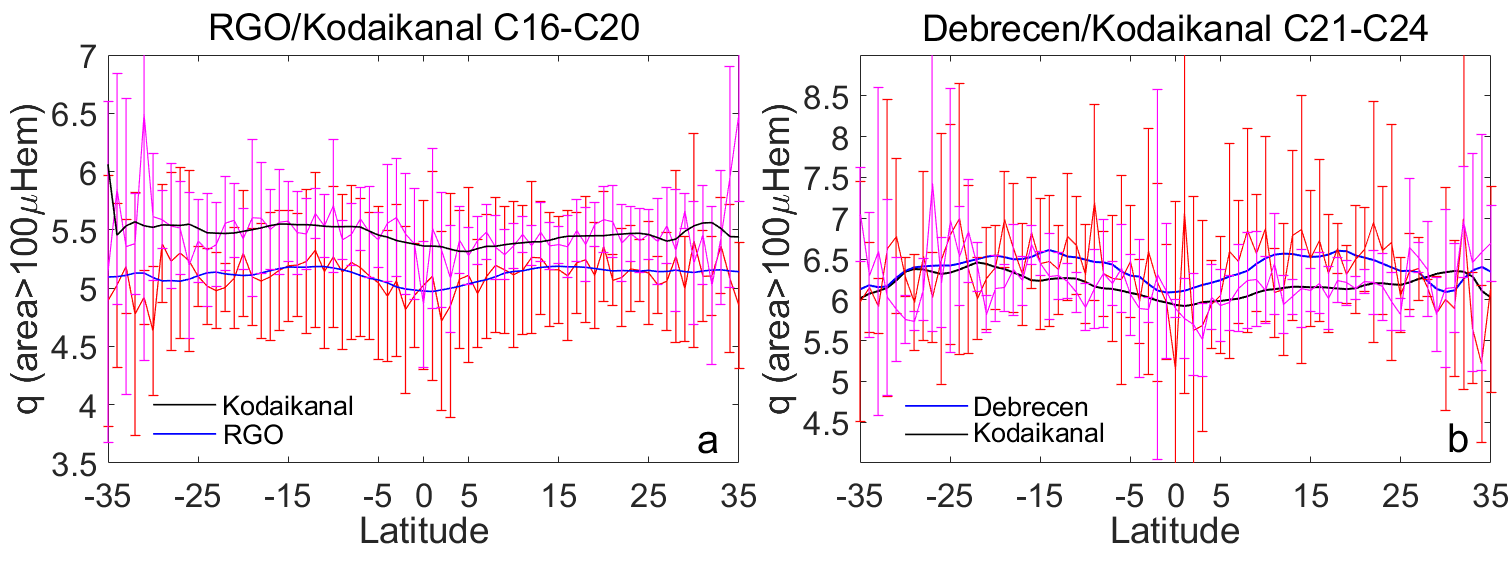}
		\caption{a) Comparison of the latitudinal penumbra-umbra ratios ($q$) for Kodaikanal and RGO Cycles 16\,--\,20 with areas larger than 100\,$\mu$Hem. b) Comparison of the latitudinal $q$-values for Debrecen and Kodaikanal Cycles 21\,--\,24 with areas larger than 100\,$\mu$Hem.}
		\label{fig:RGO_Koda_Deb_latitudinal_comparison}
\end{figure}

Figure \ref{fig:RGO_Koda_Deb_latitudinal_comparison}a shows the comparison of RGO and Kodaikanal latutudinal $q$-values for Cycles 16\,--\,20 with areas larger than 100\,$\mu$Hem. Here we also show a 13-point trapezoidal smoothing of the $q$-values. Trapezoidal smoothing is a common moving average smoothing with end points of the window having half of the weight of the inner points. The smoothed shapes of the $q$-graphs seem to be very similar with a shallow valley at the Equator. The average of the RGO $q$ is again smaller than the average of the Kodaikanal $q$, i.e. mean values (standard deviations) are 5.11 (0.15) and 5.35 (0.15) for RGO and Kodaikanal, respectively. 
Figure \ref{fig:RGO_Koda_Deb_latitudinal_comparison}b shows the comparison of Kodaikanal and Debrecen latutudinal $q$-values for Cycles 21\,--\,24 with areas larger than 100\,$\mu$Hem. It is evident that the shape of the Kodaikanal $q$-graph is very similar to that of the Kodaikanal $q$-graph in Figure \ref{fig:RGO_Koda_Deb_latitudinal_comparison}a with slightly deeper valley at the Equator in the smoothed curve. Note that the errorbars are behaving very wildly, especially for Debrecen, because the $q$-values of the odd cycles are much higher than the $q$-values of the even cycles. The Debrecen $q$-graph is, however, still deeper at the Equator having clear maxima between 10\,--\,20 degres at both hemispheres. The mean $q$-values (standard deviations) are 6.22 (0.34) and 6.36 (0.41) for Kodaikanal and Debrecen, respectively. It should, however, be noted that while it is understandable that the $q$-values are smaller near the Equator, because sunspots and sunspot groups are smaller there than around 10 to 25 degrees of latitude, the error limits also so wide, that the confidence of this result is vague.

\section{Conclusions}

We have analyzed the penumbra-umbra ratio ($q$) of sunspot latitudinal and temporal distributions for RGO SC12\,--\,SC20, Kodaikanal SC16\,--\,SC23 (also incomplete cycle 24 until june 2018) and Debrecen dataset SC21\,--\,SC24 (SC24 until the end of 2017), in order to get enough data for a solid statistics.

The analysis of $q$-values for the Kodaikanal Cycles 16\,--\,24 shows that the odd cycles have considerable more large sunspots in the ascending phase than after the maximum. On the other hand, for the even cycles the shape of the $q$-graph reminds more the shape of the sunspot cycle itself, i.e, largest sunspots are in the middle of the cycle. The mean $q$-value for even and odd cycles are 5.27 and 5.43, respectively. On the contrary, our analysis shows that the sunspots under 100\,$\mu$Hem are quite evenly distributed throughout the cycle for both the odd and even sunspot cycles. The latitudinal $q$ for large category sunspots and all sunspots seems to change such that it is lowest around the Equator and increases slightly towards higher latitudes. The latitudinal change is, however, inside the errobars and we can say that it is insignificant. Our analysis shows again that for the small sunspots ($<$100\,$\mu$Hem) the latitudinal distribution of $q$ is flat. The latitudinal average $q$ for all sunspots is same size as the temporal average q, i.e. 5.41 and 5.65 for the even and odd cycles, respectively.

The analysis of $q$ for the RGO Cycles 12\,--\,20 shows quite similar results, but now the $q$-values for the even cycles are higher than for the odd cycles. The mean values for $q$ are now 5.20 and 4.75 for the even and odd cycles, respectively. Note, that the $q$-values are moderately smaller for the RGO dataset than Kodaikanal dataset. The latitudinal analysis for RGO sunspot groups data confirms the aforementioned result, that $q$ is at lowest around Equator and increases towards higher latitudes for the large ($>$100\,$\mu$Hem) groups. Note that the errorbars for RGO data are largest, because the level of the $q$-values changes so much between cycles, especially for small category sunspot groups.

For Debrecen sunspot data we superimposed the Cycles 21\,--\,24 in the case of temporal evolution of the $q$-values. This is because of the shorter interval, i.e less cycles in the Debrecen database than for the other datasets. In this case we find an interesting phenomenon. The shape of the $q$-graph has two clear maxima and a deep valley between them for both small and large category sunspot groups. It turns out that the decrease in the $q$-values is simultaneous to the drop of the total and umbral area of the Debrecen sunspot groups. We believe that this is related to Gnevyshev gap as the deepest point of the total and umbral area locates 36\,\% from the start of the cycle \citep{Takalo_2018}. The average $q$-value for the Debrecen cycles is 5.74. 
The latitudinal $q$-values for Debrecen sunspot groups have largest variation such that they are clearly lowest (except some bad values) at the Equator of the Sun. The decrease as calculated separately for the even and odd cycles seems to be deeper than the errorbars of one standard deviation. The mean values for the even and odd Cycles 21\,--\,24 between latitudes -35\,--\,35 are 5.53 and 6.33.

\begin{figure}
	\centering
	\includegraphics[width=1.0\textwidth]{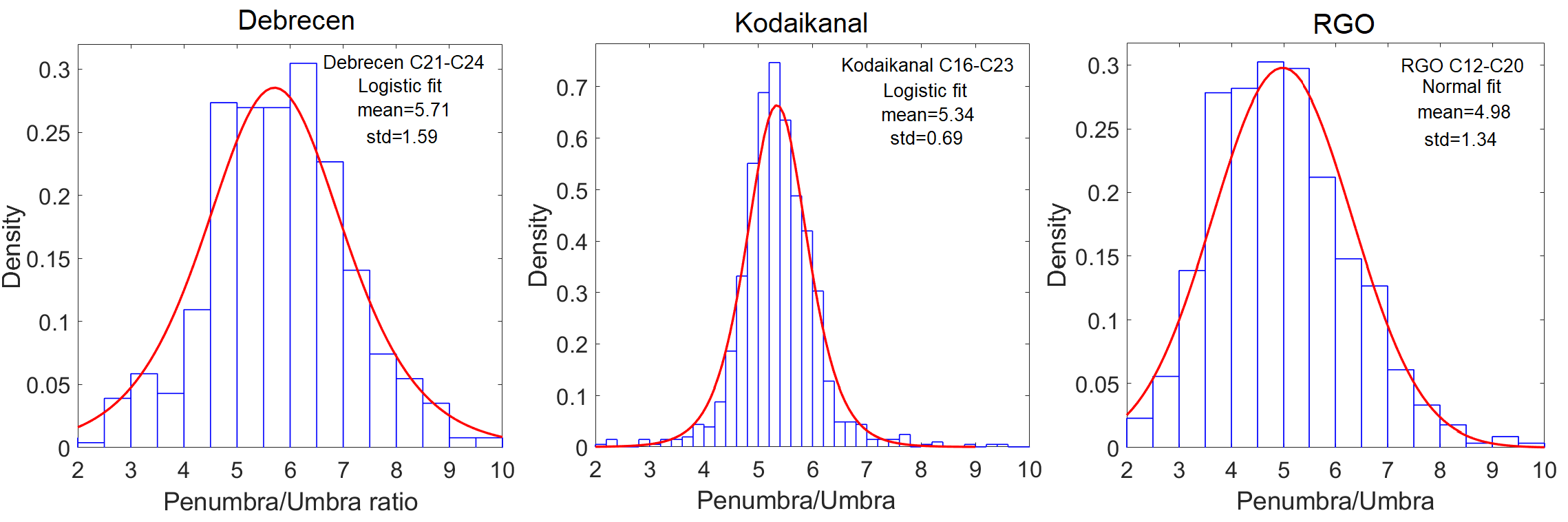}
		\caption{The relative densities of $q$-values for RGO, Kodaikanal and Debrecen databases.}
		\label{fig:RGO_Koda_Deb_densities}
\end{figure}

We have also compared the overlapping intervals of the RGO and Kodaikanal datasets for Cycles 16\,--\,20 and Kodaikanal and Debrecen datasets for Cycles 21\,--\,23. The results are consistent with the separately made analyses such that RGO has smallest $q$-values and Debrecen highest $q$-values.
The reason for the differences in $q$ is probably difficulty in separating the border between umbra and penumbra, especially for sunspot groups \citep{Steinegger_1997}. This issue is quite complicated and is not the scope of this study. Furthermore, \cite{Foukal_2014} states that the main reason why spot areas recorded using photographic or CCD observations are larger than those based on drawings seems to be that the areas of spots too small to draw are still individually measurable on good plates and CCD images. 

\begin{figure}
	\centering
	\includegraphics[width=1.0\textwidth]{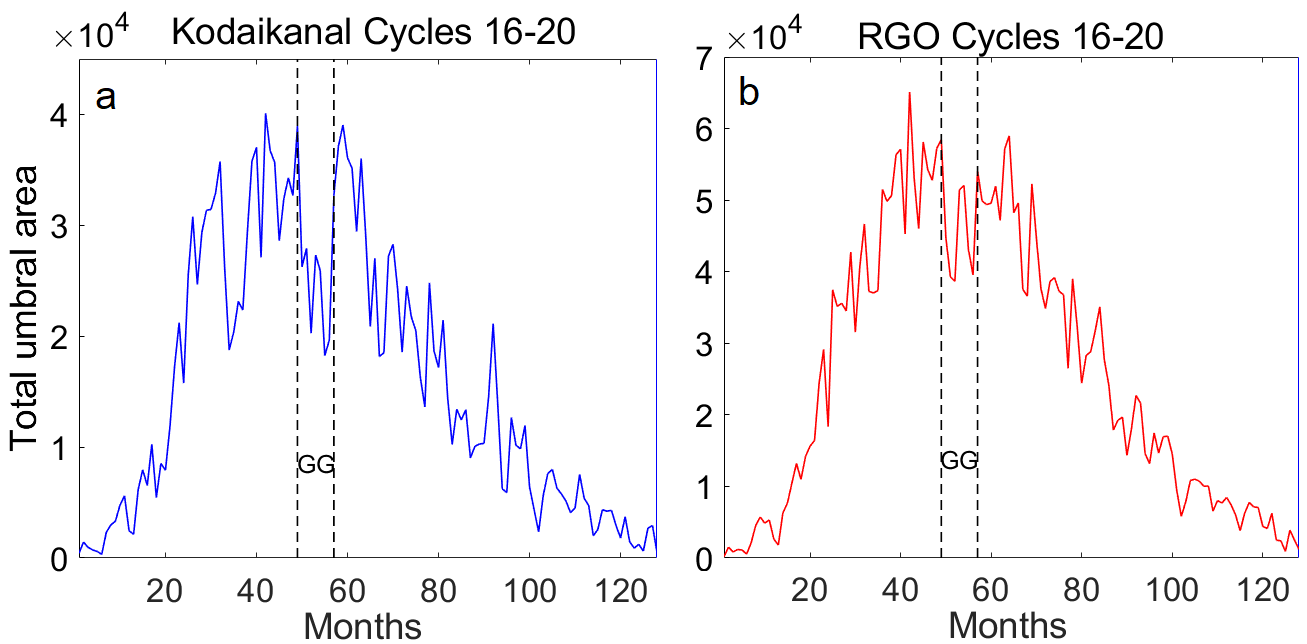}
		\caption{Total areas of a) Kodaikanal and b) RGO datasets for Solar Cycles 16\,--\,20.}
		\label{fig:RGO_Koda_total_areas}
\end{figure}

Debrecen dataset, however, seems to have smallest umbral areas, at least for large sunspots. The gradient method used in the analyses of Debrecen sunspot data is described in the article by \cite{Gyori_1998}. Debrecen dataset seems also to give most precise results in the sense that it shows also the effect of Gnevyshev gap in its temporal evolution of the $q$-values. Gnevyshev gap is somehow related to the reversal of the magnetic field of the Sun. The basic reason for the GG is not the scope of this study, and it is an open question now, but some good attempts have been presented \citep{Georgieva_2011, Karak_2018}. There have also been suggestions about a relic field in the Sun \citep{Bravo_1996, Mursula_2001, Song_2005}. If the relic field exists, it could maybe explain why GG is more intense in the even cycles than in the odd cycles \citep{Takalo_2020_3, Takalo_2020_2, Takalo_2021}. It should be noted that RGO and Kodaikanal datasets do not show the temporal decrease of the $q$-values , although they both have a clear twofold GG in their total area as a function of time for Solar Cycles 16\,--\,20 (see Figure \ref{fig:RGO_Koda_total_areas}). Note that the GG region in this case is few months later than GG of Debrecen for Solar Cycles 21\,--\,24.

Kodaikanal research group uses their own procedure in defining the border, which is based on the threshold method introduced by \cite{Otsu_1979}. Kodaikanal data seem to be most compactly located around the mean $q$-value. This is seen in Figure \ref{fig:RGO_Koda_Deb_densities}, where we show the relative densities of RGO, Kodaikanal and Debrecen databases. Kodaikanal has by far the smallest standard deviation of the $q$-distributions. Note that this standard deviation is related to the errorbars in the $q$-graphs shown earlier for each databases.  From these three datasets RGO seems to be most heterogeneous, which is seen also as a curious drop of $q$-values in the first third of the 20th century (Hathaway, 2013, see also solarscience.msfc.nasa.gov/greenwch.shtml).  Although the RGO data is one of the best sunspot area datasets, the drop in $q$-values for spots smaller than 100\,$\mu$Hem has not been seen in any other datasets so far. Thus, one must proceed with caution when comparing $q$-values between different observatories. It should also be noted, that corrections to the RGO database have been going on recently \citep{Erwin_2013}.

\begin{acknowledgements}

The RGO sunspot area data are from RGO-USAF/\newline NOAA (solarscience.msfc.nasa.gov/greenwch.shtml), Kodaikanal database \newline from github.com/bibhuraushan/KoSoDigitalArchive and Debrecen sunspot \newline datasets from  fenyi.solarobs.csfk.mta.hu/DPD/. The dates of the cycle maxima \newline were obtained from the National Geophysical Data Center (NGDC), Boulder, Colorado, USA (ftp.ngdc.noaa.gov), and the corona indices are retrieved from \newline www.ngdc.noaa.gov/stp/solar/corona.html
\end{acknowledgements}

\newpage

%%% %%%%%%%%%%%%%%%%%%%%%%%%%%%%%%%%%%%%%%%%%%%%%%%%%%%%%%%%%%%
%% Bibliography
%
% Using BibTeX
%
\bibliographystyle{spr-mp-sola}
\bibliography{references_JT_SolPhys}  
%
% Without BibTeX 
% \begin{thebibliography}{}
% \bibitem[\protect\citeauthoryear{Author}{Year}]{key}
%   <bibliographical entry>
%
% \bibitem[\protect\citeauthoryear{}{}]{}
%   
%  
% \end{thebibliography}

\end{article} 
\end{document}